\documentclass[aps,pra,showpacs,twocolumn,amsmath,amssymb]{revtex4}

\usepackage[english]{babel}
\usepackage{latexsym}
\usepackage{graphics,psfrag}
\usepackage{epsfig}
\usepackage{color}

\def\be{\begin{equation}}
\def\ee{\end{equation}}
\def\bea{\begin{eqnarray}}
\def\eea{\end{eqnarray}}
\def\bi{\begin{itemize}}
\def\ei{\end{itemize}}
\def\bin{\begin{enumerate}}
\def\ein{\end{enumerate}}
\def\la{\langle}
\def\ra{\rangle}

\begin{document}
\title{Breakdown of Anderson localization of interacting quantum bright solitons in a disorder potential}

\author{Marcin P\l{}odzie\'n}
\affiliation{
Instytut Fizyki imienia Mariana Smoluchowskiego and
Mark Kac Complex Systems Research Center, 
Uniwersytet Jagiello\'nski, ulica Reymonta 4, PL-30-059 Krak\'ow, Poland}

\author{Krzysztof Sacha}
\affiliation{
Instytut Fizyki imienia Mariana Smoluchowskiego and
Mark Kac Complex Systems Research Center, 
Uniwersytet Jagiello\'nski, ulica Reymonta 4, PL-30-059 Krak\'ow, Poland}

\date{\today}

\begin{abstract}
The center of mass of a bright soliton in a Bose-Einstein condensate may reveal Anderson localization in the presence of a weak disorder potential. We analyze the effects of interactions between two bright solitons on the Anderson localization phenomenon. Perturbation calculus shows that even very weak interactions modify localization properties of the system eigenstates. For stronger interactions, i.e. when the solitons are close to each other, the localization is totally broken. 
It implies that in order to experimentally observe the Anderson localization effects, a single bright soliton has to be prepared and excitation of soliton trains must be avoided.
\end{abstract}


\maketitle
\section{Introduction}

Fifty years ago Anderson discovered that the transport of non-interacting particles can be totally suppressed in the presence of a disorder \cite{anderson58}. Waves fail to propagate in a disorder medium, due to destructive interference, and become exponentially localized. Anderson localization possesses different properties in different dimensions \cite{gangof4}. 
In one-dimension (1D) any amount of a diagonal disorder leads to Anderson localization. Particles propagating with momentum $k$ in a disordered medium multiple scatter, undergo diffusive motion and finally localize with an exponentially decaying density profile \cite{mott61,ishii73,felderhof86,lee85,tiggelen99}.
Anderson localization is a single particle phenomenon. Interaction between particles can break the localization \cite{shepelyansky94,imry95,weinman95,flores00}. Also long-range correlations of a disorder \cite{dunlap90,bellani99,moura98,cheraghchi05} or the presence of non-linear terms in  wave-equations \cite{shepelyansky93,schulte05,shepelyansky08} can be responsible for the transition from localized to extended states. 

Ultra-cold atomic gases are ideal systems to theoretically 
and experimentally investigate quantum many-body phenomena. Anderson localization of such matter-waves was realized in a laboratory first in 1D \cite{billy08,roati08} and recently also in three-dimensions \cite{kondov2011,jendrzejewski2012}. In these experiments, in order to get rid of particle interactions, either Feshbach resonances were employed or a low atomic density limit was reached. Particle interactions, which are harmful to experiments with Anderson localization of atomic matter-waves, become vital to realize solitonic states in Bose-Einstein condensates (BEC). In ultra-cold atomic gases both dark and bright solitons were demonstrated experimentally  
\cite{burger1999,denschlag2000,khaykovich2002,strecker2002,cornish06,weller08,sengstock2008}. An analysis of quantum character of the degree of freedom that describes position of the solitons allows for invention of interesting ideas for experiments. Bright soliton scattering on a potential barrier leads to a superposition of macroscopically distinct objects \cite{weiss09,alon09,martin2012}. Interaction between two bright solitons allows for preparation of quantum entanglement of a pair of such macroscopic {\it particles} \cite{lewenstein09}. 
Bright or dark solitons can Anderson localize in the presence of an external disorder potential \cite{sacha_appa09,sacha_soliton_PRL,cord11,mochol}

In the present article we concentrate on the Anderson localization phenomena of a  pair of bright solitons in the presence of an optical speckle potential. Two bright solitons interact between each other and the interaction can lead to delocalization of their centers of mass. The soliton interaction is unusual as compared to typical particle interactions. Apart from the relative distance, the interaction potential also depends on the relative phase between solitons. 

The paper is organized as follows. In Sec.~II we describe shortly an improved version of the Bogoliubov approach that allows for the description of quantum bright solitons.
In Sec.~III we present an effective quantum Hamiltonian for two interacting solitons in a weak disorder potential. Analysis of the delocalization effects using time-independent perturbation theory and by means of numerical integration of the Schr\"odinger equation is described in Sec.~IV. We conclude in Sec.~V.

\section{Quantum bright soliton in a disorder potential}

Let us consider a Bose-Einstein condensate in an effectively 1D box of size $L$ with attractive interactions between atoms. In the mean-field description, an $N$-body quantum state is a product $\psi(z_1,t)\psi(z_2,t)\ldots\psi(z_N,t)$, where $\psi(z,t)$ fulfills the 1D Gross-Pitaevskii equation (GPE) \cite{gpe,dalfovo1999}
\begin{equation}\label{GP_1d}
i\partial_t\psi(z,t) = -\frac{1}{2}\partial_z^2\psi(z,t) - |\psi(z,t)|^2\psi(z,t).
\end{equation}
In Eq.~(\ref{GP_1d}), and in the entire paper, we have adopted the following units for energy, length and time,
\begin{align}\label{units}
 E_0 = 4m\omega_\perp^2a^2,\cr
 l_0 = \frac{\hbar}{2|a|m\omega_\perp},\cr
 t_0 = \frac{\hbar}{4a^2m\omega_\perp^2},
\end{align}
where $\omega_\perp$ denotes a transverse harmonic confinement frequency, $a$ is the atomic $s$-wave scattering length and $m$ stands for the mass of the atoms. The ground state solution of the GPE is a bright soliton that describes a localized wave-packet which evolves without changing its shape, 
\begin{align}
 \psi(z,t) &= e^{-i \mu t}\psi_0(z-r),\\
 \psi_0(z-r) &= e^{i\phi}\frac{N}{2{\rm cosh}[(z-r)N/2]},
 \label{psi0}
\end{align}
where the chemical potential $\mu=-N^2/8$. We assume that the norm of the solution equals the total particle number, i.e. $\langle\psi_0|\psi_0\rangle=N$. 
The soliton has two degrees of freedom, i.e. the center of mass position $r$ and the phase $\phi$. The bright soliton can be considered as the $N$-body bound state where an {\it ionization} of a single particle costs energy $|\mu|$ \cite{kanamoto03}.

In the Bogoliubov decription of the system
 there appear two zero modes. The first is related to the global gauge symmetry, the other originates from the translational symmetry of the $N$-particle system \cite{haus89,castin01}. Both these symmetries are broken by the classical solution (\ref{psi0}) where a specific phase of the wave-function and a concrete position of the soliton center are chosen. In order to recover the symmetries and describe degrees of freedom related to the zero modes in a non-perturbative way, we follow the Dziarmaga's approach \cite{dziarmaga}. This approach allows us to obtain a simple Hamiltonian that describes the center of mass of the soliton and its phase in the presence of a weak external potential $V(z)$
\bea
 \hat{\cal H} &\approx& \hat H_r
 +\frac{\hat P_\phi}{2m_\phi}+\hat P_{\phi}\int  V(z) \;\partial_N|\psi_0(z-r)|^2\;dz,
\cr &&
\eea
with
\bea\label{H_CM}
 \hat{H}_r &=& \frac{\hat{P}^2_r}{2 N}+\int  V(z) |\psi_0(z-r)|^2dz,
\eea
where $\psi_0$ is the solution (\ref{psi0}) and $m_\phi=-4/N$. The Hermitian operator
\bea
\hat{P}_r&=&-i\partial_r,
\eea
is the momentum operator of the soliton center of mass.
The other Hermitian operator, related to the soliton phase, is the particle number operator
\bea
\hat{P}_\phi&=&-i\partial_\phi=\hat N-N.
\eea
We may restrict ourselves to the Hilbert space with exactly $N$ particles, where $\hat P_\phi=0$, because $[\hat P_\phi,\hat {\cal H}]=0$. 

The external potential $V(z)$ also couples the center of mass degree of freedom with the Bogoliubov quasi-particles. However, this coupling has been neglected because 
if the strength of the potential is much smaller than the modulus of the chemical potential of the system, excitations of the quasi-particles can be omitted. Indeed, the lowest quasi-particle energy equals $|\mu|=N^2/8$ and for $|V(z)|\ll |\mu|$ the perturbation is not able to {\it ionize} a particle from the solitonic bound state \cite{sacha_appa09,cord11}. 

The eigenstates of the $N$-particle system can be written as follows
\be
|\Psi\ra=\chi(r)\;\xi_N(\phi)\;|0\ra_B,
\ee
where $|0\ra_B$ is the Bogoliubov quasi-particle vacuum state, $\xi_N(\phi)=1/\sqrt{2\pi}$ 
ensures that we describe the system with precisely $N$ particles, i.e. $\hat P_\phi\xi_N=0$, and $\chi(r)$ is an eigenstate of the center of mass Hamiltonian (\ref{H_CM}). The single particle density can be approximated, to the leading order in $N$, by
\be
\rho(z)=\la\hat\psi^\dagger(z)\hat\psi(z)\ra\approx \int  |\chi(r)|^2\; |\psi_0(z-r)|^2\;dr,
\ee
where $\hat\psi(z)$ is the bosonic field operator.

Equation~(\ref{H_CM}) shows that the mass center of the soliton experiences a potential which is a convolution of the original external potential with the 
soliton profile. When the external potential is a disorder potential, e.g. an optical speckle potential \cite{schulte05,billy08}, 
the center of mass of the soliton becomes Anderson localized \cite{sacha_soliton_PRL}. 
The eigenstates of the Hamiltonian (\ref{H_CM}) have a shape with an overall exponential envelope  
\be
|\chi(r)|^2\propto \exp\left(-\frac{|r-r_0|}{l_0}\right),
\ee
where $r_0$ is the mean position of the soliton and $l_0$ is the localization length. In the Born approximation and for the soliton momentum $P_r>N/(2\pi)$ 
but smaller than the inverse of the correlation length of the original speckle potential, the localization length increases exponentially 
with $P_r$ \cite{sacha_soliton_PRL}
\be
l_0\propto \exp\left(\frac{\pi P_r}{N}\right).
\ee
 Similar effects are also predicted in the dark soliton case \cite{mochol}. The position of a dark soliton becomes Anderson localized in the presence of a disorder.

\section{Quantum description of a pair of bright solitons}

Let us consider a pair of identical bright solitons where one is located at $r_1$ and the other at $r_2$. If $|r_1-r_2|$ is much greater than the soliton size, a superposition of wave-functions (\ref{psi0}) is a good approximation for the double soliton solution of the GPE (\ref{GP_1d}). In the quantum description, following a similar approach as in the single soliton case, we obtain the quantum Hamiltonian that describes the centers of mass and the phases of two independent solitons in the presence of a weak external potential
\begin{eqnarray}\label{H_2_solitons_c}
\hat H_0 & = & \sum_{i=1}^2\left[\hat H_{r_{i}} + \frac{\hat P^2_{\phi_{i}}}{2 m_{\phi_{i}}} \right. 
\cr &&
\left. +\hat P_{\phi_i}\int  V(z) \;\partial_N|\psi_0(z-r_{i})|^2\;dz \right],
\end{eqnarray}
where $\phi_{i}$'s stand for the phases of the solitons and $r_i$'s for the positions of their mass centers. The operator is: 
\be
\hat P_{\phi_i}=-i\partial_{\phi_i}=\hat N_i-N,
\ee
and $m_{\phi_i}=-4/N$. The Hamiltonian is:
\bea\label{hri}
 \hat{H}_{r_i} &=& \frac{\hat{P}^2_{r_i}}{2 N}+\int  V(z) |\psi_0(z-r_i)|^2dz,
\eea
where $\hat P_{r_i}=-i\partial_{r_i}$. The independent solitons can Anderson localize in the presence of a weak disorder potential. However, in order to analyze the localization properties of the solitons we have to take into account their mutual interaction. One soliton experiences the presence of the other one because the tails of their profiles overlap. It results in an interaction potential which drops exponentially with an increase of the solitons relative distance \cite{malomed98}
\be
\hat U  = -N^3\cos(\phi_1-\phi_2)e^{-N|r_2-r_1|/2}.
\ee
Interestingly, the character of the interaction depends on the solitons relative phase. For $\phi_1-\phi_2\approx 0$ they attract each other but if the relative phase is close to $\pi$ there are repulsive interactions. The Anderson localization of the non-interacting solitons takes place in the $r_1$ and $r_2$ space only because there is no external disorder in the $\phi_1$ and $\phi_2$ space. We will see that because the interaction potential depends on the phases,  the interaction induced delocalization is different as compared to a typical situation where only relative particle distance determines an interaction potential \cite{shepelyansky94,imry95,weinman95,flores00}.

The total particle number is a constant quantity. In order to take advantage of this constant of motion let us express the Hamiltonian of the system in terms of the average and relative phases 
\bea
\varPhi &=& \frac{\phi_1+\phi_2}{2}, 
\\
\varphi &=& \phi_2 - \phi_1, 
\eea
with the corresponding momenta 
\bea
\hat P_\varPhi &=& \hat P_{\phi_{2}} + \hat P_{\phi_{1}}=\hat N_1+\hat N_2-2N,
\\
\hat P_{\varphi}&=& \frac{\hat P_{\phi_{2}}-\hat P_{\phi_{1}}}{2}=\frac{\hat N_2-\hat N_1}{2}. 
\eea
The final Hamiltonian of two interacting bright solitons in the presence of a weak external potential reads
\be\label{htwo}
\hat H=\hat H_0 + \hat U,
\ee
where 
\bea\label{h0two}
\hat H_{0}&=& \hat H_1+\hat H_2+\frac{\hat P_\varphi^2}{2m_\varphi},
\eea
with 
\bea\label{hh1}
\hat H_1&=&\hat H_{r_{1}} -\hat P_{\varphi}\int  V(z) \;\partial_N|\psi_0(z-r_{1})|^2\;dz,
\\ \label{hh2}
\hat H_2&=& 
\hat H_{r_{2}} +\hat P_{\varphi}\int  V(z) \;\partial_N|\psi_0(z-r_{2})|^2\;dz,
\eea
and $m_\varphi=-2/N$. In the derivation of the Hamiltonian (\ref{h0two}) we have restricted it to the $2N$-particle Hilbert space where $\hat P_\varPhi=0$ that is allowed because $[\hat P_\varPhi,\hat H]=0$. Contrary to $\hat P_\varPhi$, the operator $\hat P_\varphi$ does not commute with $\hat H$ and particles can be transfered between the solitons during the time evolution of the system.

Having a quantum state $\Psi(r_1,r_2,\varphi)$, obtained from the diagonalization of the Hamiltonian (\ref{htwo}), we are able to calculate the single particle density which, in the leading order in $N$, reads
\bea\label{rhoz}
 \rho(z) &\approx& \int \left(|\psi_0(z-r_1)|^2 + |\psi_0(z-r_2)|^2 \right)
 \cr && \times \;|\Psi(r_1,r_2,\varphi)|^2\;dr_1dr_2 d\varphi.
\eea

\section{Impact of soliton interactions on Anderson localization }

In the presence of an external disorder potential $V(z)$ the center of mass of a quantum bright soliton may reveal Anderson localization. When two such solitons are prepared in the presence of a disorder, the localization effects can be destroyed due to mutual interaction between them. In the following we analyze the impact of the interaction on the Anderson localization phenomenon within the time-independent perturbation approach and by means of numerical integration of the Schr\"odinger equation generated by the Hamiltonian (\ref{htwo}).

\begin{figure}
\centering
\includegraphics*[width=0.95\linewidth]{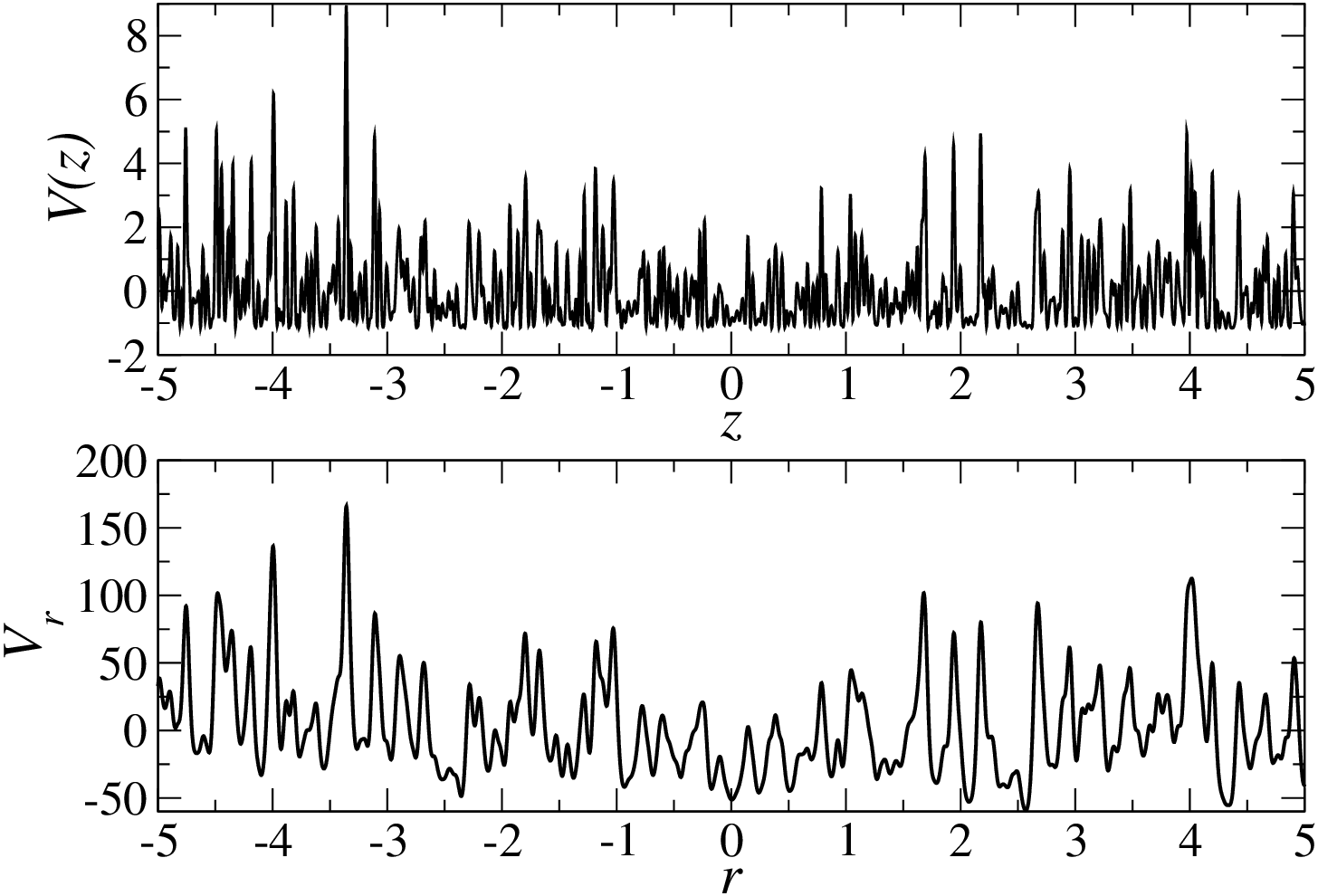}
\caption{Top panel: an example of an optical speckle potential $V(z)$ for correlation length $\sigma_0=0.01$ and strength $V_0=1.125$. Bottom panel: disorder potential experienced by a center of mass of a bright soliton, i.e. after convolution with the soliton profile,  $V_r = \int V(z)|\psi_0(z-r)|^2$, for $N=60$.
}
\label{one}
\end{figure}

We will consider a finite system of length $L$ and focus on a disorder potential corresponding to an optical speckle potential \cite{schulte05,billy08}. Such a potential can be created by shining laser light on the so-called diffusive plate. In the far field the light intensity fluctuates in space which is experienced by atoms as an external disorder potential. The potential is characterized by zero mean value $\overline{V(z)}=0$, where the overbar denotes an ensemble average over disorder realizations, standard deviation
$V_0=\left[\overline{V(z)^2}\right]^{1/2}$ and autocorrelation function $\overline{V(z')V(z'+z)}=V_0^2\frac{\sin^2(z/\sigma_0)}{(z/\sigma_0)^2}$ where $\sigma_0$ is the correlation length of the disorder. An example of an optical speckle potential is plotted in Fig.~\ref{one}. Interestingly, properties of a speckle potential can be easily modified. Especially, one can create a disorder where Anderson localization operates as a band-pass filter that allows for the realization of a matter-wave analog of an optical random laser \cite{arl11,aspect11}.
\subsection{Time independent perturbation approach}

In this section we investigate eigenstates of the Hamiltonian (\ref{htwo}) treating the interaction potential
 $\hat{U}$ as a small perturbation. The energies and eigenstates of the unperturbed system ($\hat U=0$) read 
\begin{eqnarray}
E^{(0)}_{n_{1} n_{2} n_{\varphi}} &=& E_{n_{1}n_\varphi} + E_{n_{2}n_\varphi} + E_{n_{\varphi}}, \cr
\Psi^{(0)}_{n_{1} n_{2} n_{\varphi}}(r_1,r_2,\varphi)&=&\chi_{n_{1}n_\varphi}(r_1)\chi_{n_{2}n_\varphi}(r_2)\xi_{n_{\varphi}}(\varphi), \cr &&
\end{eqnarray}
where $\xi_{n_\varphi}(\varphi)$ is an eigenstate of the $\hat P_\varphi$ operator related to the eigenvalue $n_\varphi$. The energy $E_{n_\varphi}$ is an eigenvalue of the $\hat P_\varphi^2/(2m_\varphi)$ operator. We assume that the numbers of particles that form each of the two solitons is equal to $N$. Thus, the unperturbed system is characterized by $n_\varphi=0$ and $\xi_0(\varphi)=1/\sqrt{2\pi}$. The states $\chi_{n_{1}n_\varphi}(r_1)$ and $\chi_{n_{2}n_\varphi}(r_2)$  are eigenstates of the Hamiltonians (\ref{hh1}) and (\ref{hh2}), respectively, and  $E_{n_{1}n_\varphi}$ and $E_{n_{2}n_\varphi}$ are the corresponding eigenenergies with $n_\varphi=0$. These eigenstates are Anderson localized in the presence of the disorder.

The first order energy correction vanishes,
\begin{equation}
 E_{n_{1} n_{2} 0}^{(1)} = \langle\Psi_{n_{1} n_{2} 0}^{(0)}|\hat{U}|\Psi_{n_{1} n_{2} 0}^{(0)}\rangle=0.
\end{equation}
This is because $ \hat{U} \sim \cos\varphi$. The only states $\Psi_{n_{1}'n_{2}'n_{\phi}'}^{(0)}$ coupled by the interaction potential to $\Psi_{n_{1} n_{2} 0}^{(0)}$ correspond to $n_\varphi'=\pm 1$. The lowest order energy correction is of the second order $E_{n_{1} n_{2} 0}^{(2)}$. In the lowest non-vanishing order the system eigenstates read
\bea\label{psi_correction}
\Psi_{n_1 n_2 0}&\approx&\Psi_{n_1 n_2 0}^{(0)}+ \sum_{{n_1',n_2'}} \alpha_{n_1'n_2'1}\Psi^{(0)}_{n'_{1} n'_{2} 1} \\
 &+& \sum_{{n_1'',n_2''}}\alpha_{n_1''n_2''-1}\Psi^{(0)}_{n''_{1} n''_{2} -1},
\eea
where 
\be
\alpha_{n_1'n_2'1}=\frac{\langle\Psi_{n_1'n_2' 1}|\hat{U}|\Psi_{n_1n_2 0}\rangle}{E^{(0)}_{n_1 n_2 0} - E^{(0)}_{n_1' n_2' 1}},
\ee
\be
\alpha_{n_1''n_2'' -1}=\frac{\langle\Psi_{n_1''n_2''-1}|\hat{U}|\Psi_{n_1n_2 0}\rangle}{E^{(0)}_{n_1 n_2 0} - E^{(0)}_{n_1'' n_2'' -1}}.
\ee
Reduced probability density for finding the first soliton localized at $r_1$ is 
\bea \label{kap}
 \kappa_1(r_1) &=& \int \left|\Psi_{n_{1} n_{2} 0}(r_1,r_2,\varphi)\right|^2d\varphi dr_2 \cr
&=& 
|\chi_{n_{1}0}(r_1)|^2 + \sum_{n'_{2}}\left|\sum_{n'_{1}}\alpha_{n'_{1} n'_{2} 1} \chi_{n_{1}'1}(r_1)\right|^2 \cr
&&+ \sum_{n''_{2}}\left|\sum_{n''_{1}}\alpha_{n''_{1} n''_{2} -1} \chi_{n_{1}''-1}(r_1)\right|^2, \cr &&
\eea 
and the probability $\kappa_2(r_2)$ for finding the second soliton at $r_2$ can be calculated analogously.

As an example we choose parameters of an optical speckle potential like in Ref.~\cite{billy08} where Anderson localization of a BEC has been realized. 
The length of the system, in the units (\ref{units}), $L=28$~(i.e. 4mm), the transverse harmonic confinement corresponds to $\omega_\perp=2\pi\times 70$ Hz 
and the correlation length of the disorder $\sigma_0=0.01$ (i.e. 1.32$\mu$m). We assume that each of the two solitons consists of $N=60$ $^{85}$Rb atoms with 
the modified $s$-wave scattering length by means of a Feshbach resonance, i.e. $a=-5.77$~nm. The optical speckle potential is created by blue detuned light with 
an intensity which results in the disorder strength  $V_0=1.125$ that is 400 times smaller than the chemical potential of the system.

\begin{figure}
\centering
\includegraphics*[width=0.95\linewidth]{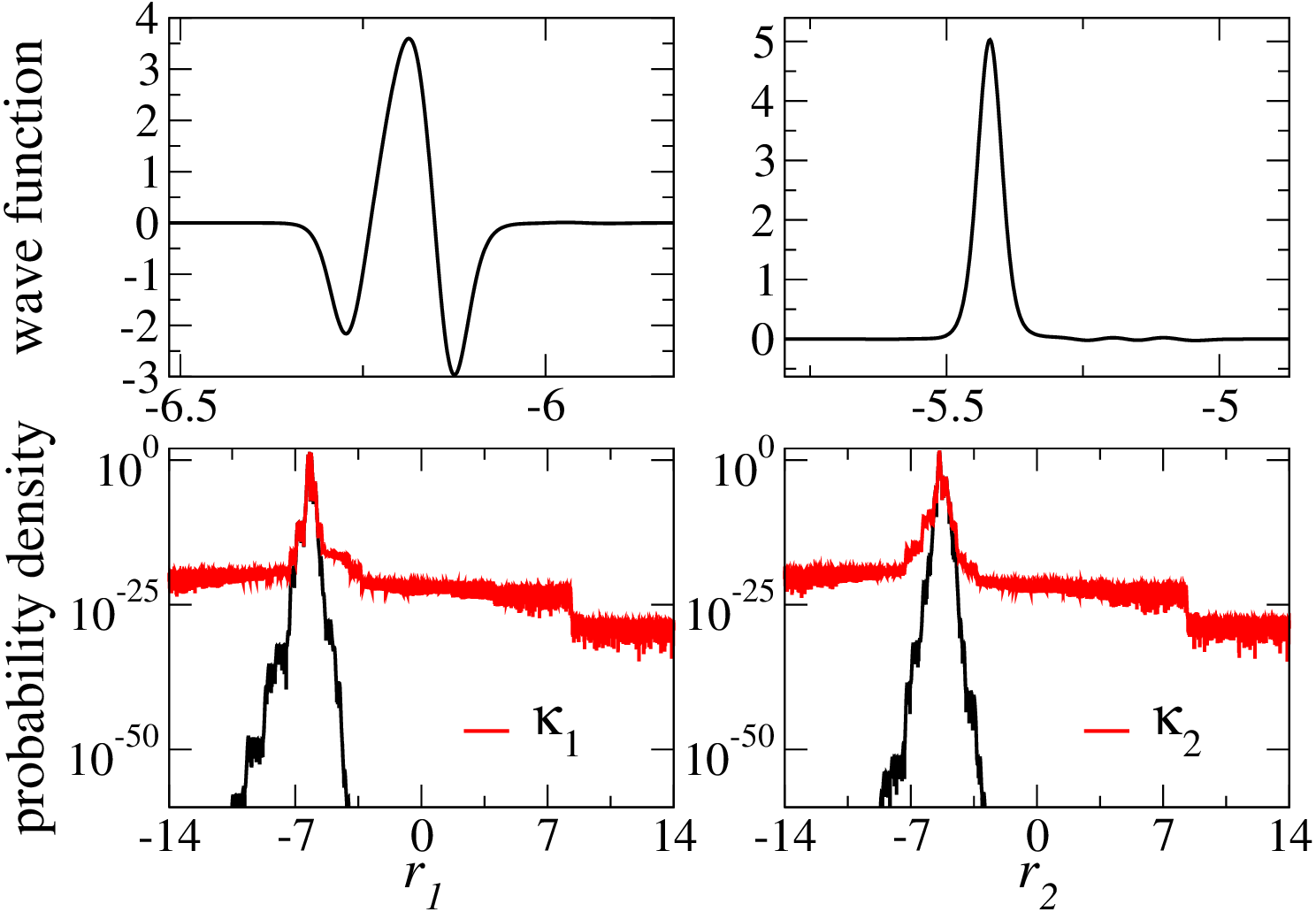}
\caption{(Color online) 
Top panels: eigenstates, $\chi_{n_{1}0}(r_1)$ and $\chi_{n_{2}0}(r_2)$, of the Hamiltonians (\ref{hh1}) and (\ref{hh2}) corresponding to the eigenvalues $E_{n_10} = 9.93$ and $E_{n_20} = 14.52$. Localization lengths are $l_{n_10} \approx 0.020 $, $l_{n_20} \approx 0.026$. Bottom panels: probability densities $|\chi_{n_{1}0}(r_1)|^2$ and $|\chi_{n_{2}0}(r_2)|^2$ (black, bottom lines) and the reduced probability densities for centers of mass of solitons $\kappa_1(r_1)$ and $\kappa_2(r_2)$ (red, upper lines) obtained within the perturbation approach. The correlation length of the disorder potential is $\sigma_0 = 0.01$ and its  strength $V_0 = 1.125$. Total number of particles in the system $2N=120$.
}
\label{two}
\end{figure}

\begin{figure}
\centering
\includegraphics*[width=0.95\linewidth]{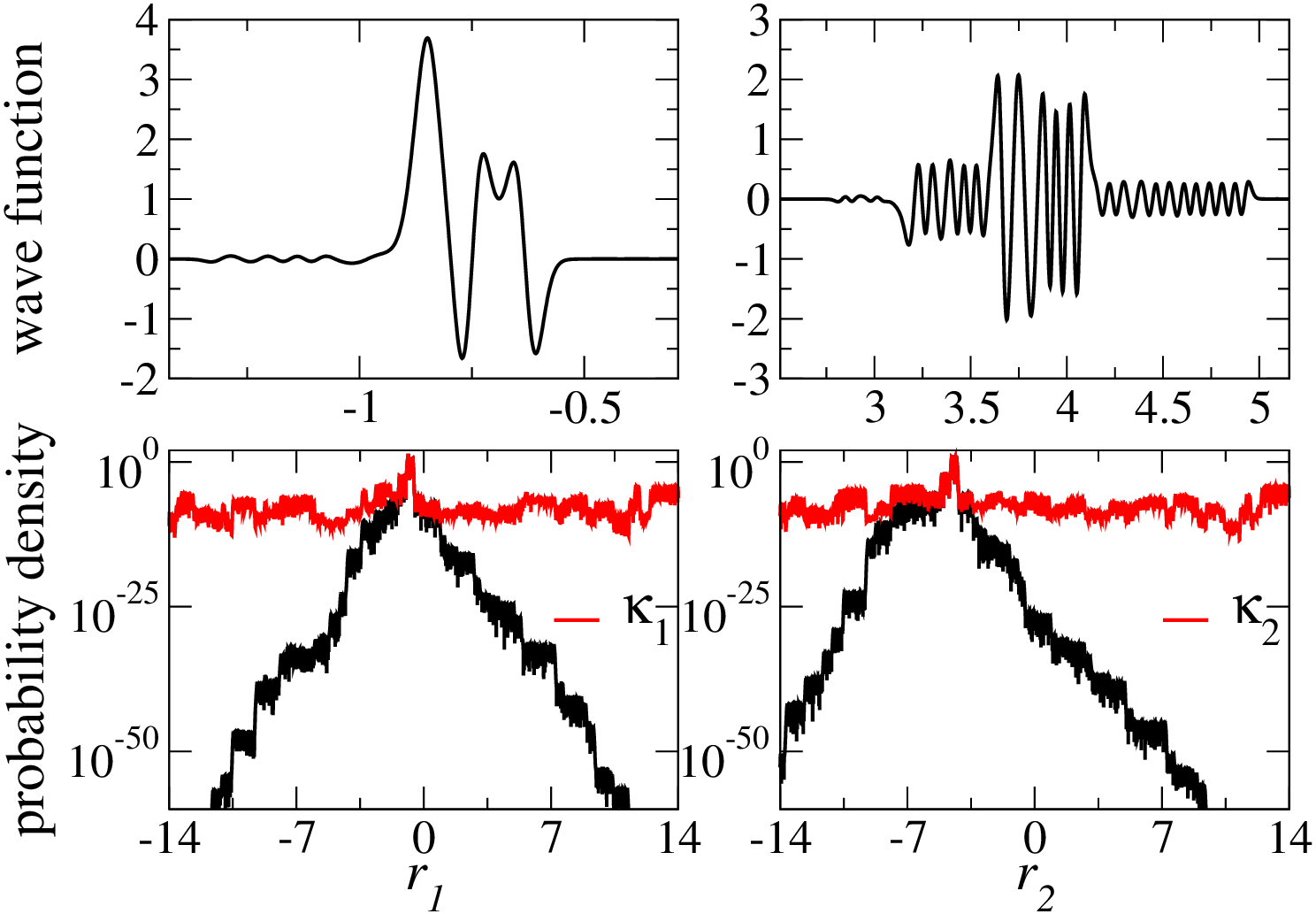}
\caption{(Color online) 
The same as in Fig. 2 but for $E_{n_10} = 42$ and $E_{n_20} = 46.6$.
Localization lengths of the eigenstates $\chi_{n_{1}0}(r_1)$ and $\chi_{n_{2}0}(r_2)$ are $l_{n_10} \approx 0.088 $ and $l_{n_20} \approx 0.095$, respectively.
}
\label{three}
\end{figure}

Diagonalization of the Hamiltonians (\ref{hh1})-(\ref{hh2}) allows us to obtain unperturbed Anderson localized eigenstates $\chi_{n_10}(r_1)$ and $\chi_{n_20}(r_2)$. We have chosen two pairs of these eigenstates for further analysis. The first pair is related to eigenenergies $E_{n_10}=9.93$ and $E_{n_20}=14.52$ and the estimated localization lengths $l_{n_10}\approx0.020$ and $l_{n_20}\approx0.026$. They are relatively strongly localized on a scale comparable to the soliton size $2/N=0.03$. The other pair is related to $E_{n_10}=42$ and $E_{n_20}=46.6$, and 
$l_{n_10}\approx0.088$ and $l_{n_20}\approx0.095$, respectively. Both pairs of the eigenstates are shown in Fig.~\ref{two} and Fig.~\ref{three}. The relative distance between the solitons in each pair is much larger than the soliton size which implies that the soliton interaction is extremely weak and the perturbation approach is justified. In the considered cases, the second order energy corrections $E^{(2)}_{n_1n_20}$ are 7 orders of magnitude smaller than energy gap between $E^{(0)}_{n_1n_20}$ and the nearest neighboring level.

Figures~\ref{two}-\ref{three} also show the reduced probability densities $\kappa_i(r_i)$, Eq.~(\ref{kap}). Delocalization effects of the solitons are hardly visible when we look at the densities in linear scale because the soliton interactions are very weak. However, logarithmic plots indicate that characteristic exponential profiles are not present. 
The mechanism of the breakdown of Anderson localization in a finite system, considered here, is the following. The interaction potential couples an unperturbed state $\Psi^{(0)}_{n_1n_20}$ to states $\Psi^{(0)}_{n_1'n_2'1}=\chi_{n_1'1}\chi_{n_2'1}\xi_{1}$ and $\Psi^{(0)}_{n_1''n_2''-1}=\chi_{n_1''-1}\chi_{n_2''-1}\xi_{- 1}$ where wave-functions $\chi_{n_i'1}(r_i)$ and $\chi_{n_i''-1}(r_i)$ reveal greater localization lengths the higher the corresponding eigenenergies $E_{n_i'1}$ and $E_{n_i''-1}$ are. If the localization lengths start to be comparable to the system size $L$ the wave-functions $\chi_{n_i' 1}(r_i)$ and $\chi_{n_i''-1}(r_i)$ become delocalized, extended states. Contributions of the extended states in (\ref{psi_correction}) are responsible for the breakdown of the exponential decay that can be observed in Figs.~\ref{two}-\ref{three}. In the unperturbed state $\Psi^{(0)}_{n_1n_20}$, the numbers of particles forming each soliton is well defined and equal to $N=60$. Corrections, $\alpha_{n_1'n_2'1}\Psi^{(0)}_{n'_{1} n'_{2} 1}$ and $\alpha_{n_1''n_2''-1}\Psi^{(0)}_{n''_{1} n''_{2} -1}$, to the unperturbed state indicate that the delocalization is accompanied by transfers of a particle from one soliton to the other. 
By decreasing the distance between the unperturbed, localized solitons we increase the interaction energy and breakdown of exponential profiles start closer to the localization centers. If the unperturbed solitons are located far away from each other it may happen that the corrections to the unperturbed state are too small to disturb exponential localization in a finite system. This is in contrast to a more typical situation where an interaction potential depends only on the relative distance between particles. 
In fact, suppose for a moment that $\hat U$ is only a function of $|r_1-r_2|$, then the degenerate unperturbed states $\Psi^{(0)}_{n_1n_20}$ and $\Psi^{(0)}_{n_2n_10}$ are coupled by $\hat U$. Diagonalization of $\hat U$ within the degenerate subspace results on a linear combination of these states. Thus, even if $\hat U\rightarrow 0$, each soliton is described by $\kappa_i(r_i)$ which is a sum of the densities $|\chi_{n_10}(r_i)|^2$ and $|\chi_{n_20}(r_i)|^2$ that are localized at two different positions. In the case considered in this publication such a mechanism is not present and the breakdown of the Anderson localization is an higher order effect.

\begin{figure}
\centering
\includegraphics*[width=0.95\linewidth]{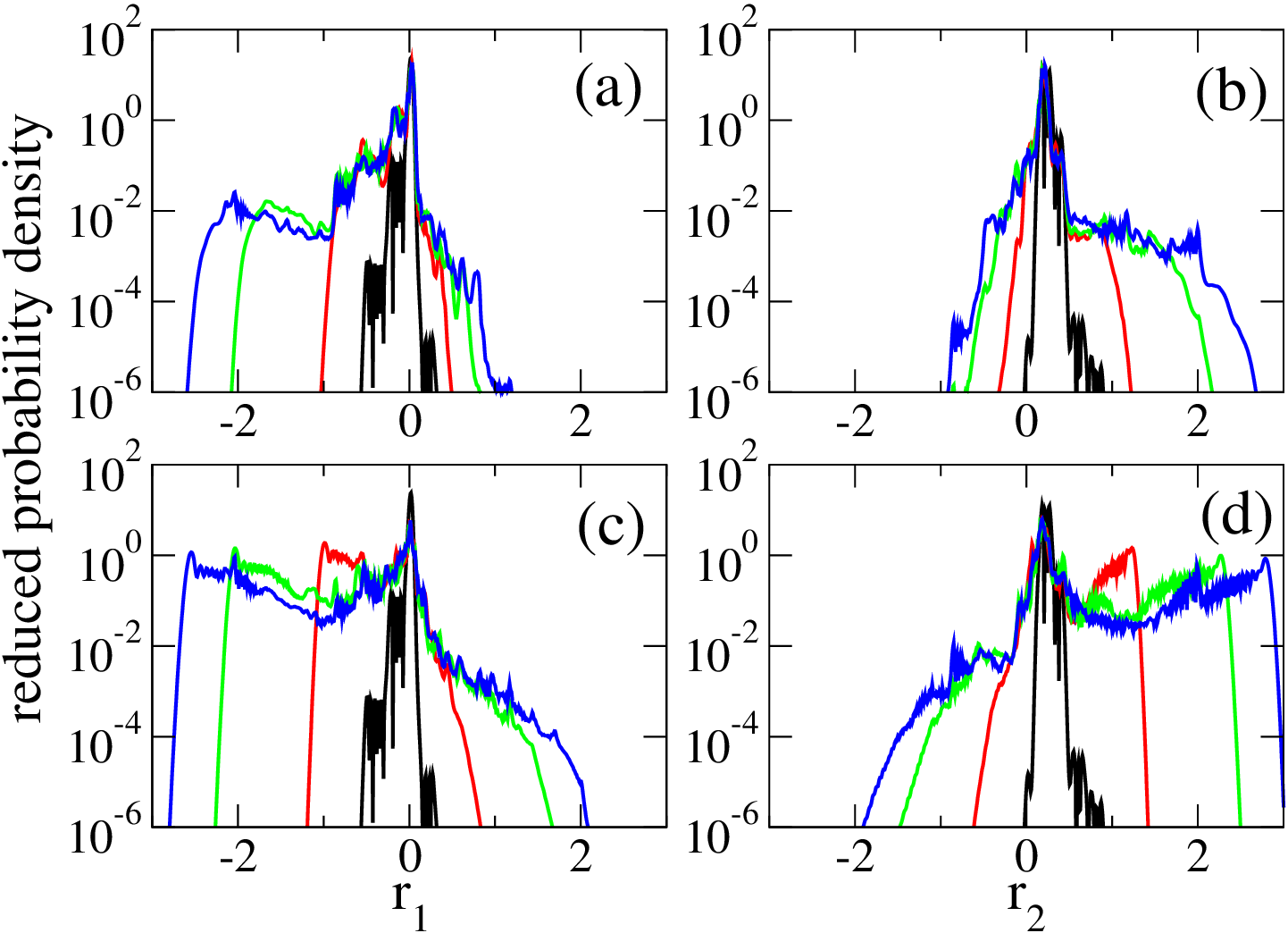}
\caption{(Color online) 
Reduced probability densities $\kappa_i(r_i)$, Eq.~(\ref{kap}), for different moments of time. In each panel, from the bottom to the top, the curves are related to: $t=0$ (black), $t = 0.5$ (red), $t = 1$ (green) and $t = 1.25$ (blue). The unit of time is $t_0=29$s, see Eq.~(\ref{units}). Panels (a) and (b) show the reduced probability densities $\kappa_1(r_1)$ and $\kappa_2(r_2)$, respectively, and are related to the case where the solitons attract each other, i.e. the initial state $\xi(\varphi)$ in Eq.~(\ref{psi_init}) is chosen as the ground state of the effective Hamiltonian (\ref{hphieff}). Panels (c) and (d) are related to the case where the solitons initially repel each other, i.e. the initial state $\xi(\varphi)$ is the ground state of the effective Hamiltonian (\ref{hphieff}) shifted so that the maximal density is located at $\varphi=\pi$. The initial states $\chi_{n_10}(r_1)$ and $\chi_{n_20}(r_2)$ in Eq.~(\ref{psi_init}) correspond to $E_{n_10} = 10.44 $ and $E_{n_20} = 12.88$ and their localization length are $l_{n_1 0 } \approx 0.029$ and $l_{n_2 0} \approx  0.032$. Size of the system is $L=8$, the correlation length of the disorder potential $\sigma_0 = 0.01$ and its strength $V_0 = 1.125$. Total number of particles in the system $2N=120$.
}
\label{four}
\end{figure}

\begin{figure}
\centering
\includegraphics*[width=0.95\linewidth]{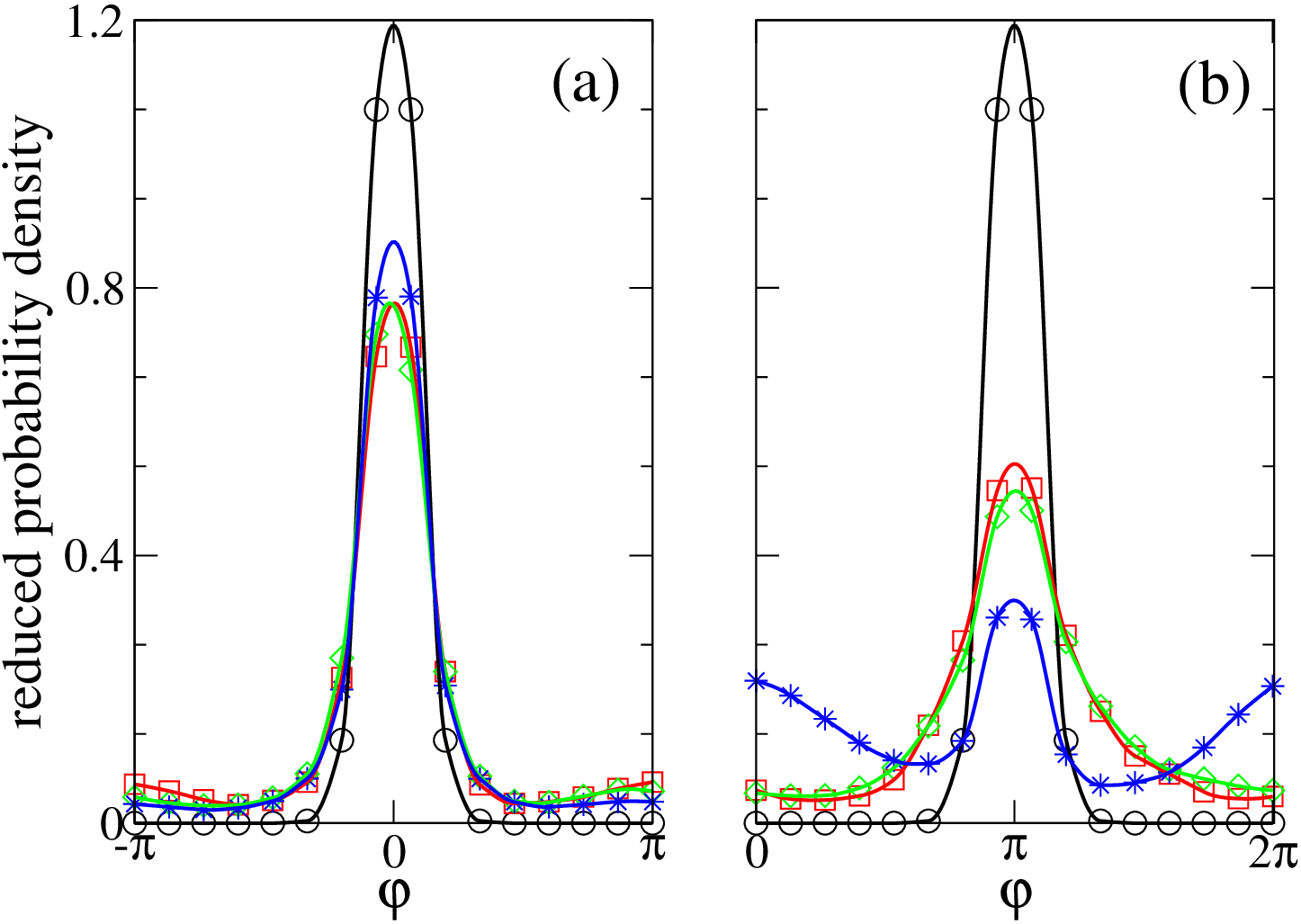}
\caption{(Color online) 
Reduced probability density $\kappa_\xi(\varphi)$, Eq.~(\ref{kap_ksi}), for different moments of time. In both panels black circles are related to $t=0$, red squares to $t = 0.5$, green diamonds to $t = 1$ and blue stars to $t = 1.25$. Panel (a) corresponds to the results presented in Fig.~\ref{four}a-b, i.e. to the case when the solitons attract each other. Panel (b) is related to the data shown in Fig.~\ref{four}c-d, i.e. to the case when the solitons initially repel each other. Note that the range of the horizontal axis in the panels is different, i.e. $(-\pi,\pi)$ in (a) and $(0,2\pi)$ in (b).
}
\label{five}
\end{figure}

\begin{figure}
\centering
\includegraphics*[width=0.95\linewidth]{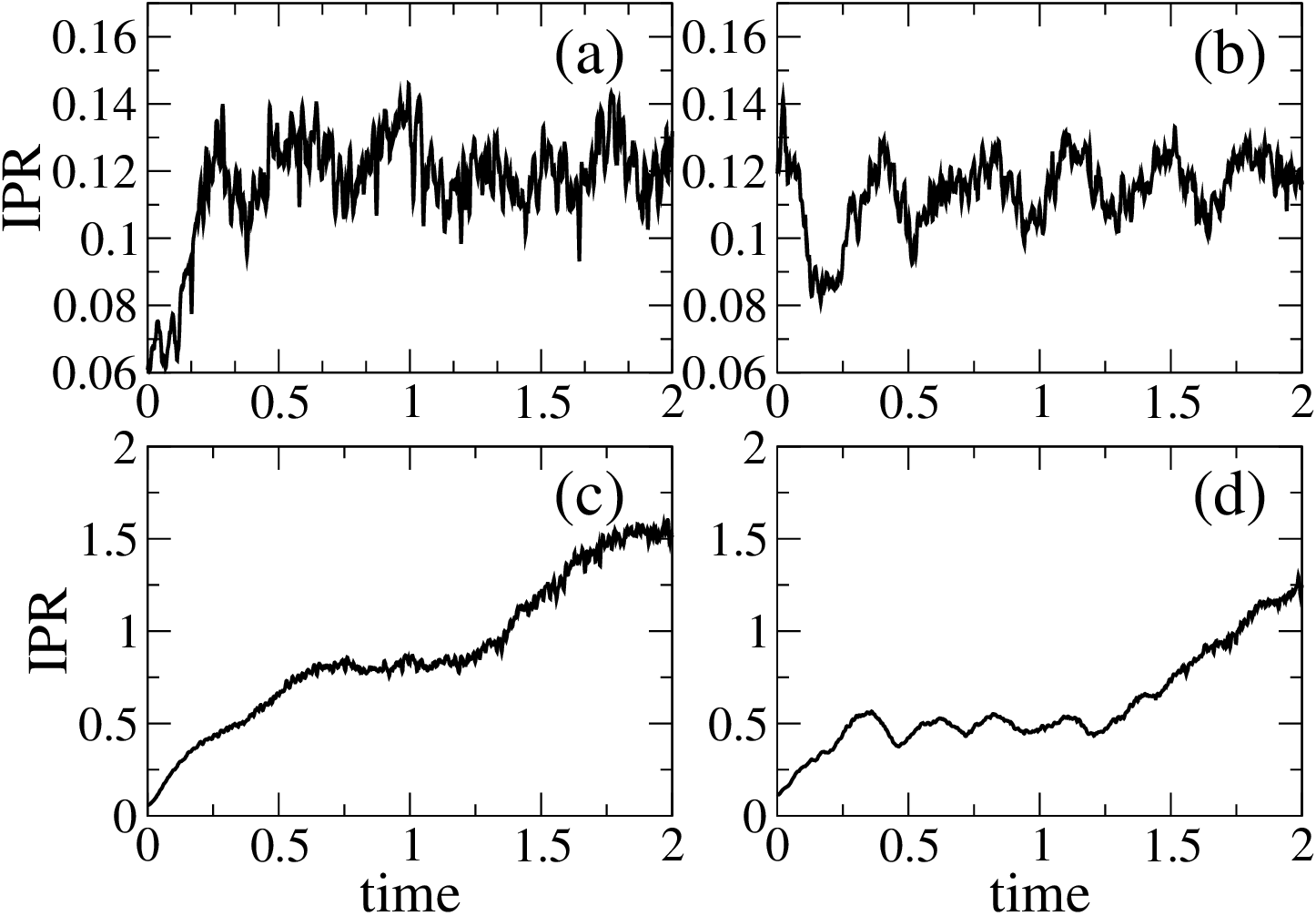}
\caption{   Evolution of the IPR, see Eq.~(\ref{ipreq}). Panels (a)-(b) correspond to the results presented in Fig.~\ref{four}a-b, i.e. to the case when the solitons attract each other. Panels (c)-(d) are related to the data shown in Fig.~\ref{four}c-d, i.e. to the case when the solitons initially repel each other. 
}
\label{iprplot}
\end{figure}

\begin{figure}
\centering
\includegraphics*[width=0.95\linewidth]{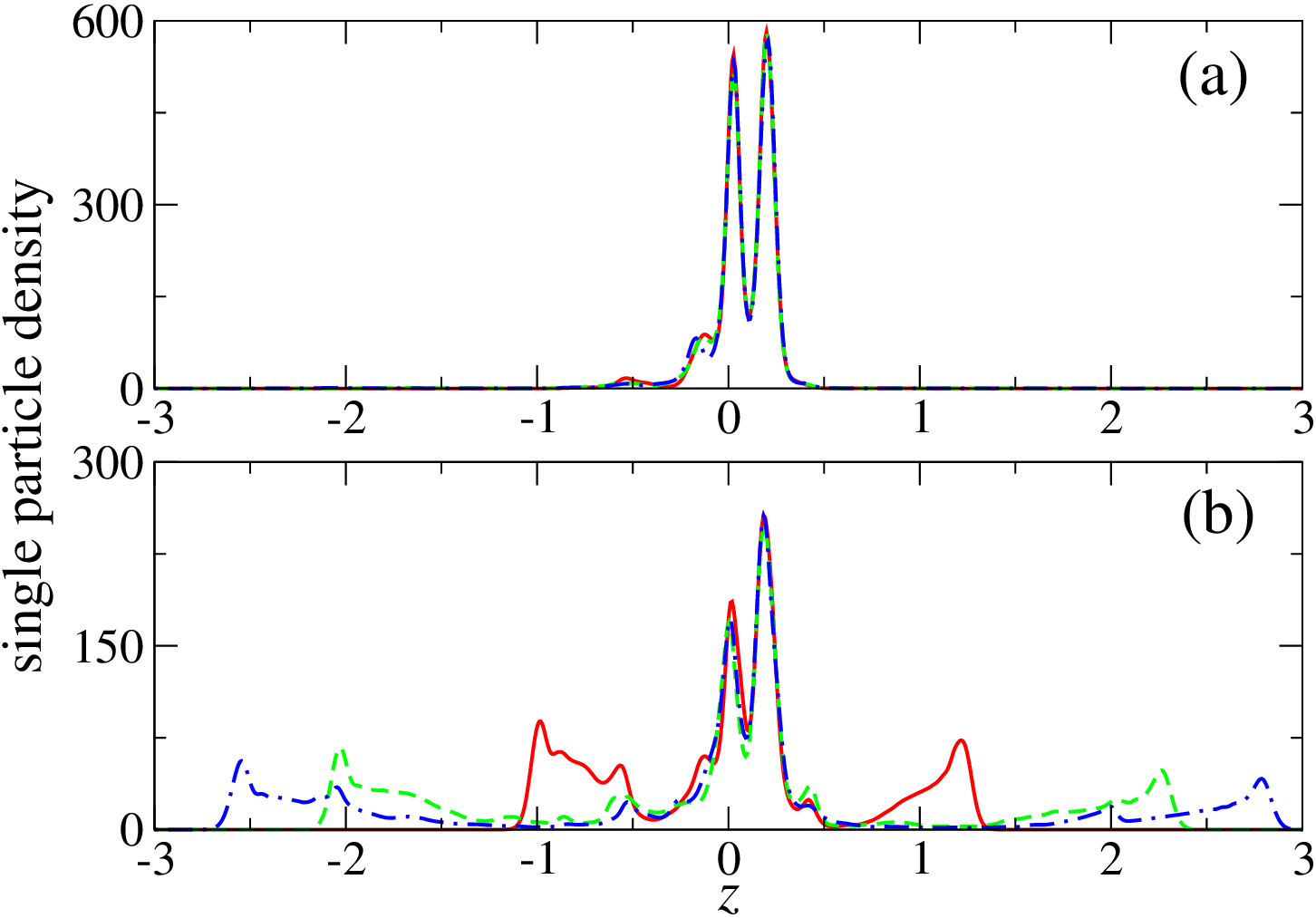}
\caption{(Color online) 
Single particle density $\rho(z)$, Eq.~(\ref{rhoz}), for different moments of time. In both panels red solid line is related to $t = 0.5$, green dashed line to $t = 1$, blue dotted/dashed line to $t = 1.25$. Panel (a) corresponds to the results presented in Fig.~\ref{four}a-b, i.e. to the case when the solitons attract each other. Panel (b) is related to the data shown in Fig.~\ref{four}c-d, i.e. to the case when the solitons initially repel each other. 
}
\label{six}
\end{figure}

\subsection{Beyond perturbation approach: numerical integration of the Schr\"odinger equation}

The interaction induced delocalization effects, described in the previous section, are very weak. This is due to the fact that by employing the perturbation approach we have to restrict ourselves to situations where unperturbed solitons are localized far away from each other. In this section we will see that if the solitons are initially localized sufficiently close to each other the interactions can totally destroy the localization of solitons.

In the present section we show results of numerical integration of the Schr\"odinger equation corresponding to the Hamiltonian (\ref{htwo}). We assume that initially the system is prepared in the following state
\begin{equation}\label{psi_init}
 \Psi(r_1,r_2,\varphi) = \chi_{n_10}(r_1) \chi_{n_20}(r_2) \xi(\varphi),
\end{equation}
where $\chi_{n_i0}$ are eigenstates of the Hamiltonians (\ref{hh1})-(\ref{hh2}) corresponding to eigenenergies $E_{n_10}=10.44$ and $E_{n_20}=12.88$. They are Anderson localized around $r_1\approx 0.18$ and $r_2\approx 0$ with the localization lengths $l_{n_10}\approx 0.029$ and $l_{n_20}\approx0.032$. The parameters of the disorder potential are the same as in the previous section, however, the size of the system has been reduced to $L=8$. In numerical integration of the Schr\"odinger equation the space $(r_1,r_2,\varphi)$ has been discretized and we have adopted absorbing boundary conditions at the ends of the $r_i$ axes and periodic boundary conditions in the $\varphi$ degree of freedom. The numerical integration is time consuming because in order to reproduce the correlation length of the disorder, the number of points in the $r_i$ spaces have to be very big. On the other hand, number of points corresponding to the $\varphi$ degree of freedom can be chosen to be quite small. 

Time evolution of the relative phase $\varphi$ is much faster than the evolution of the centers of mass of the solitons  because the corresponding {\it mass} parameters are very different, i.e. $m_\varphi=-2/N$ while the soliton masses are equal $N$. We choose as an initial state $\xi(\varphi)$ the ground state of an effective Hamiltonian
\be\label{hphieff}
\hat H_{\varphi,\rm eff}=\frac{\hat P_\varphi^2}{2m_\varphi}+\hat U_{\rm eff}(\varphi) + \hat{P}_\varphi \beta,
\ee
where 
\bea
\label{betaparam}
\beta &=& \int dzdr \left[|\chi_{n_20}(r)|^2-|\chi_{n_10}(r)|^2\right]\times
\cr && \times V(z)\;\partial_N |\psi_0(z-r)|^2. \cr
\hat U_{\rm eff}(\varphi)&=&\int dr_1dr_2 \;|\chi_{n_10}(r_1)|^2 \;|\chi_{n_20}(r_2)|^2 \times \cr && \times \hat U(r_1,r_2,\varphi)\propto -\cos\varphi. 
\eea
Such a ground state is strongly peaked around $\varphi=0$ which implies that initially the solitons attract each other. In time-evolution of the state (\ref{psi_init}) we may expect that the reduced probability density 
\be\label{kap_ksi}
\kappa_\xi(\varphi)=\int \left|\Psi(r_1,r_2,\varphi)\right|^2 dr_1dr_2,
\ee
follows the probability density of an instantaneous ground state of the Hamiltonian (\ref{hphieff}) where $\hat U_{\rm eff}$ and $\beta$ are obtained by substituting in Eq.~(\ref{betaparam}) $|\chi_{n_i0}(r_i)|^2$ by time-evolving $\kappa_i(r_i)$.

Figure~\ref{four}a-b shows time-evolution of the reduced probability densities $\kappa_i(r_i)$. Interaction between solitons is responsible for the breakdown of the initial Anderson localization of the solitons. The delocalization effects are much stronger than those analyzed in the previous section. Plots of the reduced probability density $\kappa_\xi(\varphi)$ presented in Fig.~\ref{five}a, indicate that while $\kappa_\xi(\varphi)$ does not precisely follow the density of the instantaneous ground state of (\ref{hphieff}),
the initial degree of coherence between the solitons is practically preserved in the time-evolution. The density $\kappa_\xi(\varphi)$ is concentrated around zero, i.e. the relative phase between the solitons $\varphi\approx 0$ and consequently  the solitons always attract each other. 

In Figs.~\ref{four}-\ref{five} we also show the evolution of the reduced probability densities in the case when the solitons initially repel each other. That is, as an initial state $\xi(\varphi)$ in (\ref{psi_init}) we have chosen the ground state of the effective Hamiltonian (\ref{hphieff}) but this state has been shifted so that the density maximum is not localized around $\varphi=0$ but around $\varphi=\pi$. In the present case the delocalization of the solitons is stronger than in the case when the solitons attract each other. Moreover, initial repulsion of the solitons weakens in time, i.e. the probability density $\kappa_\xi(\varphi)$ becomes nearly uniformly distributed with small maxima around both $\varphi=\pi$ and $\varphi=0$.

In order to analyze the delocalization of the solitons quantitatively we present time evolution of the inverse participation ratio (IPR) in Fig.~\ref{iprplot}. The inverse participation ratio,
\be
{\rm IPR}=\left[\int \kappa_i^2(r)dr\right]^{-1},
\label{ipreq}
\ee
is the length scale on which the center of mass of  a soliton is localized \cite{haake}. While the delocalization of the solitons in the case of the attractive soliton interactions is not very big (Fig.~\ref{iprplot}a-b), the solitons that initially repel each other reveal substantial delocalization on the time scale presented in Fig.~\ref{iprplot}. Indeed, the IPR shown in Fig.~\ref{iprplot}c-d is comparable to the size of the system $L=8$.

In Ref.~\cite{alon11}, the many-body evolution of a BEC, where all atoms occupy a double soliton solution of the GPE, has been analyzed by numerical simulations. The authors consider the cases where the relative phase between the solitons is precisely defined and it is either 0 or $\pi$. In the both cases they observe a loss of phase coherence between the solitons which leads to dramatic fragmentation of the condensate. Our results show that if the solitons attract each other and the initial state for the phase degree of freedom is chosen properly, the phase coherence changes very little in the timescale needed for delocalization of the soliton centers of mass. Indeed, if we choose as an initial state for the phase degree of freedom the ground state of the effective Hamiltonian (\ref{hphieff}), the probability density (\ref{kap_ksi}) remains localized at the bottom of the effective potential $\hat U_{\rm eff}$, Eq.~(\ref{betaparam}), during time evolution. 

If a pair of bright solitons is prepared in a laboratory, each density measurement is expected to reveal two solitons \cite{castin01,delande12} located at positions $r_i$ with probabilities $\kappa_i(r_i)$. Averaging atomic density over many realizations of the same experiment we obtain a density profile which can be compared to the single particle density $\rho(z)$, see Eq.~(\ref{rhoz}). In Fig.~\ref{six} we show the time evolution of $\rho(z)$.  
Even on a linear scale the delocalization effects can be clearly visible, especially in the case when the solitons initially repel each other.

Experimental realization of the Anderson localization of a single soliton and observation of the delocalization effects in the case of a pair of interacting solitons requires the absence of decoherence effects. The most dangerous are atomic losses which localize centers of mass of the solitons and prevent interference phenomena needed in the Anderson localization process. In the present paper we concentrate on $^{85}$Rb atoms in the vicinity of the Feshbach resonance at magnetic field of 155 G. The scattering length assumed in our analysis corresponds to the magnetic field of about 169 G. For this magnetic field the two- and three-body loss rates measured experimentally are $10^{-14}$ cm$^3/s$ and $10^{-27}$ cm$^6/s$, respectively, and the background loss time is
450~s \cite{roberts}. Assuming that the density profile of the atomic cloud 
is a product of the 1D soliton density and the density of the harmonic oscillator ground state in the transverse directions, the resulting lifetime is 160~s, i.e. 5.5 in the units (\ref{units}). This seems to be sufficient time to see the delocalization effects analyzed in the present publication.

\section{Conclusion}

We have considered quantum bright solitons in the presence of an external disorder potential. When a single soliton is placed in the disorder, Anderson localization of its center of mass is predicted by \cite{sacha_soliton_PRL}. If two solitons are present the mutual interaction between them can be responsible for the breakdown of the localization. We have analyzed this phenomenon within the perturbation approach
and by means of numerical integration of the Schr\"odinger equation. 
The perturbation approach shows that exponential tails of the probability densities for the centers of mass of the solitons disappear due to coupling of an unperturbed eigenstate to high energy delocalized states that is induced by the solitons interaction. The interaction induced delocalization is different from the typical situation \cite{shepelyansky94,imry95,weinman95,flores00} because apart from the relative distance between the solitons, the interaction potential depends also on the relative phase between them. 
In the numerical integration, where we can afford stronger interactions than in the perturbation analysis, we observe that the localization of the solitons can be totally broken.

The  results obtained in the present publication can be verified in experiments. They also indicate that in order to observe experimentally the Anderson localization of solitons, a single soliton has to be excited and preparation of soliton trains should be avoided.

\section*{Acknowledgments}

This work is supported by Polish National Science Centre under projects 
DEC-2011/01/N/ST2/00418 (MP) and DEC-2011/01/B/ST3/00512 (KS).



\begin{thebibliography}{99}  

\bibitem{anderson58}
P. W. Anderson, Phys. Rev. {\bf 109}, 1492 (1958).

\bibitem{gangof4}
P. W. Anderson, D. C. Licciardello, T. V. Ramakrishnan, E. Abrahams, Phys. Rev. Lett. {\bf 42}, 673 (1979).

\bibitem{mott61}
N. F. Mott, W. D. Twose, Adv. Phys. {\bf 10}, 107 (1961).

\bibitem{ishii73}
K. Ishii, Suppl. Prog. Theor. Phys. {\bf 53}, 77 (1973).

\bibitem{felderhof86}
B. U. Felderhof, J. Stat. Phys. {\bf 43}, 267 (1986).

\bibitem{lee85}
P. A. Lee and T.V. Ramakrishnan, Rev. Mod. Phys. {\bf 57}, 287 (1985).

\bibitem{tiggelen99}
B. van Tiggelen, in {\it Diffuse Waves in Complex Media},
edited by J.-P. Fouque, NATO Advanced Study Institutes,
Ser. C, Vol. 531 (Kluwer, Dordrecht, 1999).

\bibitem{shepelyansky94}
D. L. Shepelyansky, Phys. Rev. Lett. {\bf 73}, 2607 (1994).

\bibitem{imry95}
Y. Imry, Europhys. Lett. {\bf 30}, 405 (1995).

\bibitem{weinman95}
D. Weinmann, A. M\"{u}ller-Groeling, J.-L. Pichard, K. Frahm, Phys. Rev. Lett. {\bf 75}, 1598 (1995)

\bibitem{flores00}
J. C. Flores, Phys. Rev. B {\bf 62}, 33 (2000).

\bibitem{dunlap90}
David H. Dunlap, H-L. Wu, and Philip W. Phillips, Phys. Rev. Lett. {\bf 65}, 88 (1990).

\bibitem{bellani99}
V. Bellani, E. Diez, R. Hey, L. Toni, L. Tarricone, G. B. Parravicini, F. Dom\'{\i}­nguez-Adame, and R. G\'{o}mez-Alcal\`{a}, Phys. Rev. Lett. {\bf 82}, 2159 (1999).

\bibitem{moura98}
F.A.B.F. de Moura, M. L. Lyra, Phys. Rev. Lett {\bf 81}, 3735 (1998).

\bibitem{cheraghchi05}
H. Cheraghchi, S. M. Fazeli, K. Esfarjani, Phys. Rev. B {\bf 72}, 174207 (2005).

\bibitem{shepelyansky93}
D. L. Shepelyansky, Phys. Rev. Lett. {\bf 70}, 1787 (1993).

\bibitem{schulte05}
J. T. Schulte, S. Drenkelforth, J. Kruse, W. Ertmer, J. Arlt, K. Sacha, J. Zakrzewski, and M. Lewenstein, Phys. Rev. Lett. {\bf 95}, 170411 (2005).

\bibitem{shepelyansky08}
A. S. Pikovsky, D. L. Shepelyansky, Phys. Rev. Lett. {\bf 100}, 094101 (2008).

\bibitem{billy08}
J. Billy, V. Josse, Z. Zuo, A. Bernard, B. Hambrecht, P. Lugan, D. Cl\'{e}ment, L. Sanchez-Palencia, P. Bouyer, A. Aspect, Nature {\bf 453}, 891 (2008).

\bibitem{roati08}
G. Roati, C. D'Errico, L. Fallani, M. Fattori, C. Fort, M. Zaccanti, G. Modugno, M. Modugno, M. Inguscio, Nature {\bf 453}, 895 (2008).

\bibitem{kondov2011}
S. S. Kondov, W. R. McGehee, J. J. Zirbel, B. DeMarco, Science {\bf 334}, 66 (2011).

\bibitem{jendrzejewski2012}
F. Jendrzejewski, A. Bernard, K. M\"{u}ller, P. Cheinet, V. Josse, M. Piraud, L. PezzÃ©, L. Sanchez-Palencia, A. Aspect, P. Bouyer, Nature Physics {\bf 8}, 398 (2012).

\bibitem{burger1999}
S. Burger, K. Bongs, S. Dettmer, W. Ertmer, K. Sengstock, A. Sanpera, G. V. Shlyapnikov, M. Lewenstein,
Phys. Rev. Lett. {\bf 83}, 5198 (1999).

\bibitem{denschlag2000}
J. Denschlag, J. E. Simsarian, D. L. Feder, Charles W. Clark, L. A. Collins, J. Cubizolles, L. Deng, E. W. Hagley, K. Helmerson, W. P. Reinhardt, S. L. Rolston, B. I. Schneider and W. D. Phillips, Science {\bf 287}, 97 (2000).

\bibitem{khaykovich2002}
L. Khaykovich, F. Schreck, G. Ferrari, T. Bourdel, J. Cubizolles, L. D. Carr, Y. Castin, C. Salomon,
Science {\bf 296}, 1290 (2002).

\bibitem{strecker2002}
K. E. Strecker, G. B. Partridge, A. G. Truscott, R. G. Hulet, Nature {\bf 417}, 150 (2002).

\bibitem{cornish06}
S. L. Cornish, S. T. Thompson, C. E. Wieman, Phys. Rev. Lett. {\bf 96}, 170401 (2006).

\bibitem{weller08}
A. Weller, J. P. Ronzheimer, C. Gross, J. Esteve, M. K. Oberthaler, Phys. Rev. Lett. {\bf 101}, 130401 (2008).

\bibitem{sengstock2008}
C. Becker, S. Stellmer, P. Soltan-Panahi, S. D\"{o}rscher, M. Baumert, E. M. Richter, J. Kronj\"{a}ger, K. Bongs, K. Sengstock, Nature Physics {\bf 4}, 496 (2008) 

\bibitem{weiss09}
C. Weiss and Y. Castin, Phys. Rev. Lett. {\bf 102}, 010403 (2009).

\bibitem{alon09}
A. I. Streltsov, O. E. Alon, and L. S. Cederbaum, Phys. Rev. A {\bf 80}, 043616
(2009).

\bibitem{martin2012}
A. D. Martin, J. Ruostekoski, New J. Phys. {\bf 14} 043040 (2012).

\bibitem{lewenstein09}
M. Lewenstein and B. A. Malomed, New. J. Phys. {\bf 11}, 113014 (2009).

\bibitem{sacha_soliton_PRL}
K. Sacha, C. A. M\"{u}ller, D. Delande, J. Zakrzewski, Phys. Rev. Lett. {\bf 103}, 210402 (2009).

\bibitem{sacha_appa09}
K. Sacha, D. Delande, J. Zakrzewski, Acta Physica Polonica A {\bf 116}, 772 (2009).

\bibitem{cord11}
C. A. M\"{u}ller, Appl. Phys. B {\bf 102}, 459 (2011).

\bibitem{mochol}
M. Mochol, M. P\l{}odzie\'n, K. Sacha,
Phys. Rev. A {\bf 85}, 023627 (2012).

\bibitem{gpe}
L.P. Pitaevskii, Sov. Phys. JETP {\bf 13}, 451 (1961); E.P.
Gross, Nuovo Cimento {\bf 20}, 454 (1961).

\bibitem{dalfovo1999}
 F. Dalfovo, S. Giorgini, L. P. Pitaevskii, and S. Stringari, Rev. Mod.
Phys. {\bf 71}, 463 (1999).

\bibitem{kanamoto03}
R. Kanamoto, H. Saito, and M. Ueda, Phys. Rev. A {\bf 67}, 013608 (2003).

\bibitem{haus89}
Y. Lai, H. A. Haus, Phys. Rev. A {\bf 40}, 844 (1989).

\bibitem{castin01}
Y. Castin, in 'Coherent atomic matter waves', 
Lecture Notes of Les Houches Summer School, p.1-136, edited by R. Kaiser, C. Westbrook, and F. David, EDP Sciences and Springer-Verlag (2001).

\bibitem{delande12}
D. Delande, K. Sacha, M. P\l{}odzie\'n, S. K. Avazbaev, J. Zakrzewski,
arXiv:1207.2001.

\bibitem{dziarmaga}
J. Dziarmaga, Phys. Rev. A, {\bf 70}, 063616 (2004).

\bibitem{malomed98}
B. A. Malomed, Phys. Rev. E {\bf 58}, R864 (1998).

\bibitem{arl11}
M. P\l{}odzie\'n, K. Sacha, Phys. Rev. A {\bf 84}, 023624 (2011).

\bibitem{aspect11}
M. Piraud, A. Aspect, L. Sanchez-Palencia, arXiv:1104.2314.


\bibitem{haake}
F. Haake, {\it Quantum Signatures of Chaos}, Springer-Verlag Berlin Heidelberg 2010.

\bibitem{alon11}
A. I. Streltsov, O. E. Alon, L. S. Cederbaum, Phys. Rev. Lett. {\bf 106}, 240401 (2011).

\bibitem{roberts}
J. L. Roberts, N. R. Claussen, L. S. Cornish, and C. E. Wieman, Phys. Rev. Lett. {\bf 85}, 728 (2000).


\end{thebibliography}
\end{document}